\documentclass[a4paper]{jpconf}
\usepackage{citesort}
\usepackage{bm}
\usepackage{amssymb}
\usepackage{epsfig}
\usepackage{graphicx}
\usepackage{color}
\usepackage{subfigure}
\newcommand{\be}{\begin{equation}}
\newcommand{\ee}{\end{equation}}
\newcommand{\beq}{\begin{equation}}
\newcommand{\eeq}{\end{equation}}

\newcommand\bea{\begin{eqnarray}}
\newcommand\eea{\end{eqnarray}}

\newcommand{\apjl}{The Astrophysical Journal}
\newcommand{\apj}{The Astrophysical Journal}
\newcommand{\mnras}{Monthly Notices of the Royal Astronomical Society}
\newcommand{\pre}{Physical Review E}

\begin{document}
\title{Calculations in the theory of tearing instability}
\author{Stanislav Boldyrev$^{1,2}$ and Nuno\ F.\ Loureiro$^{3}$}
\address{${~}^1$Department of Physics, University of Wisconsin, Madison, WI 53706, USA}
\address{${~}^2$Space Science Institute, Boulder, Colorado 80301, USA}
\address{${~}^3$Plasma Science and Fusion Center, Massachusetts Institute of Technology, Cambridge MA 02139, USA}
\ead{boldyrev@wisc.edu}

%\author{Stanislav Boldyrev}
%\affiliation{Department of Physics, University of Wisconsin at Madison, Madison, WI 53706, USA}
%\affiliation{Space Science Institute, Boulder, Colorado 80301, USA}
%\author{Nuno\ F.\ Loureiro}
%\affiliation{Plasma Science and Fusion Center, Massachusetts Institute of Technology, Cambridge MA 02139, USA}

\date{\today}

\begin{abstract}
Recent studies have suggested that the tearing instability may play a significant role in magnetic turbulence. In this work we review the theory of the magnetohydrodynamic tearing instability in the general case of an arbitrary tearing parameter, which is relevant for applications in turbulence. We discuss a detailed derivation of the results for the standard Harris profile and accompany it by the derivation of the results for a  lesser known sine-shaped profile. We devote special attention to the exact solution of the inner equation, which is the central result in the theory of tearing instability. We also briefly discuss the influence of shear flows on tearing instability in magnetic structures. Our presentation is self-contained; we expect it to be accessible to researchers in plasma turbulence who are not experts in magnetic reconnection.      
\end{abstract}
%\pacs{52.35.Ra, 52.35.Vd, 52.30.Cv}
%\keywords{magnetic fields --- magnetohydrodynamics --- turbulence}
%\maketitle

%***********************************************************************************

\section{Introduction}
Numerical simulations and analytic models have suggested that magnetic plasma turbulence tends to form anisotropic, sheet-like current structures at small scales \cite{matthaeus_turbulent_1986,biskamp2003,servidio_magnetic_2009,servidio_magnetic_2011,wan2013,zhdankin_etal2013,tobias2013,zhdankin_etal2014,davidson2017,chen2017}. Such structures are not necessarily associated with the dissipation scale of turbulence. Rather, a hierarchy of sheet-like turbulent eddies is formed throughout the whole inertial interval \cite{boldyrev2005,boldyrev_spectrum_2006,chen_3D_2012,chandran_intermittency_2015,mallet_measures_2016}. Recently, it has been realized that given large enough Reynolds number, such anisotropic structures  may become unstable to the tearing mode at scales well above the Kolmogorov-like dissipation scale \cite{loureiro2017,mallet2017,boldyrev_2017,loureiro2017a,mallet2017a,comisso2018}.    
The Reynolds numbers for which such effects become significant are very large ($Re\gtrsim 10^6$), so their definitive study is beyond the capabilities of modern computers. Nevertheless, the rapid progress in in situ measurements of space plasma brings interest to small scales of magnetic turbulence, where such effects may be observed, e.g.,~\cite{vech2018}. It is, therefore, highly desirable to develop an understanding of the linear tearing theory in magnetic profiles such as those one might expect to find throughout the inertial range of turbulence, but not necessarily those associated with dissipative current sheets.

This brings attention to the two facets of the theory of tearing instability that are not traditionally covered in textbooks on magnetohydrodynamics or plasma physics. One is the theory of reconnection beyond the well-known Furth, Killeen \& Rosenbluth regime of small tearing parameter~\cite{FKR}. This regime assumes limited anisotropy of a reconnecting magnetic profile, so it is not applicable to very anisotropic tearing modes relevant for our study. The other is the theory of tearing instability for the magnetic profiles that are different from the canonical $\tanh$-like Harris profile~\cite{harris_1962}. Such a profile assumes that the reconnecting magnetic field is uniform in space except for the region where it reverses its direction. This is, arguably, not a general situation encountered in turbulence, where the magnetic fields strength varies in space on similar scales both inside and outside the reconnection region. Different magnetic profiles may lead to different scalings of the corresponding tearing growth rates, e.g.,~\cite{boldyrev_2017,loureiro2017a,walker2018,loureiro2018,pucci2018}. To the best of our knowledge, there are currently no texts methodically covering these aspects of tearing instability. Rather, various relevant analytical results are scattered over the literature, e.g.,~\cite{coppi1966,coppi_resistive_1976,abc1978,porcelli_viscous_1987,loureiro_instability_2007}.    

In this work we review the derivation of the standard Harris-profile tearing mode, and  accompany it by a parallel derivation  of the results for the sine-shaped profile. We devote special attention to the discussion of the exact solution of the inner equation, for which we use a method different from those previously adopted in the literature \cite{coppi1966,coppi_resistive_1976,abc1978,porcelli_viscous_1987,loureiro_instability_2007}. Although our work is mostly devoted to the tearing instability in magnetic profiles not accompanied by velocity fields, in the end of our presentation we discuss to what extent shear flows, typically present within turbulent eddies, can modify the results. The goal of our work is to give a self-contained presentation of the theory of tearing effects that are most relevant for applications to turbulence. We believe it will be useful for researchers in turbulence who are not necessarily experts in reconnection.

\section{Equations for the tearing mode}
We assume that the background uniform magnetic field is in the $z$ direction, and the current sheet thickness, $a$, and length, $l$,  are measured in the field-perpendicular plane. The current sheets are strongly anisotropic, $a\ll l$. We denote the reconnecting magnetic field, that is, the variation of the magnetic field across the current sheet, as~$B$. Such structures can be created in turbulence if their life times are comparable to the Alfv\'enic time $\tau_A\sim l/V_{A}$, where $V_{A}$ is the Alfv\'en speed associated with~$B$. They are formed at all scales, the thinner the structure the more anisotropic it becomes. A theory describing a hierarchy of such magnetic fluctuations, or turbulent eddies, in MHD turbulence suggests that their anisotropy increases as their scale decreases, $a/l\propto a^{1/4}$ \cite{boldyrev2005,boldyrev_spectrum_2006,mason_cb06,mason2012,perez_etal2012,chandran_15}. 

It has been proposed that at small enough scales the tearing instability of very anisotropic eddies can compete with their Alfv\'enic dynamics.\footnote{These ideas stem from the observation that if current sheets were allowed to have arbitrarily large aspect ratios, they would be tearing unstable at rates that diverge when the Lundquist number tends to infinity~\cite{pucci_reconnection_2014,uzdensky_loureiro2016,tenerani2016}.} This means that below a certain critical scale the tearing time should become comparable to $\tau_A$, so that the turbulence is mediated by tearing instability \cite{loureiro2017,mallet2017,boldyrev_2017}. Such a picture has received some numerical and observational support \cite{walker2018,dong2018,vech2018}. The theory of tearing instability required to describe strongly anisotropic current sheets, goes beyond the simplified FKR theory and generally requires one to analyze  structures that are different from the Harris-type current sheets. 

It should be acknowledged that the first analysis of the tearing instability in structures formed by MHD turbulence dates back to 1990~\cite{carbone1990}.\footnote{We were not aware of this important early work at the time when our previous studies~\cite{loureiro2017,boldyrev_2017} were published.} That analysis was based on the Iroshnikov-Kraichnan  model of MHD turbulence~\cite{iroshnikov_turbulence_1963,kraichnan_inertial_1965} that treats turbulent fluctuations as essentially isotropic (that is, characterized by a single scale) weakly interacting Alfv\'en wave packets. 
 The anisotropy of turbulent fluctuations has therefore not been quantified in \cite{carbone1990}. Moreover, their model assumed the presence of a significant velocity shear in the current layer and adopted the tearing-mode growth rate calculated in the shear-modified FKR regime \cite{hofmann1975,dobrowolny1983,einaudi1986}. As a result, the approach of \cite{carbone1990} was qualitatively different from that of the recent studies~\cite{loureiro2017,mallet2017,boldyrev_2017}.   

In our discussion we do not impose any limitations on the anisotropy of current structures, that is, we assume them to be anisotropic enough to accommodate the fastest growing tearing mode; this assumption is consistent with the model of MHD turbulence \cite{boldyrev_spectrum_2006} adopted in \cite{loureiro2017,boldyrev_2017}. We do, however, make several  important simplifications. First, we assume that the background magnetic field has only one component, $B_y(x)$. An optional uniform guide field in $z$-direction may also be present; it has no effect on the problem. Second, in most of the work we assume that the configuration is static, that is, there is no background flows. In section~\ref{shear_flow} we briefly discuss the effects of a shear flow, where, similarly to the magnetic field, the shearing velocity field is assumed to have only one component, $v_y(x)$. (We refer the reader to, e.g., \cite{chen_morrison1990,tolman2018,loureiro2013} for a broader discussion of the possible effects of background flows and strong outflows). Finally, our analysis is limited to the MHD framework.

\begin{figure}[t]
\begin{center}
\begin{minipage}{4in}
\includegraphics[width=\columnwidth]{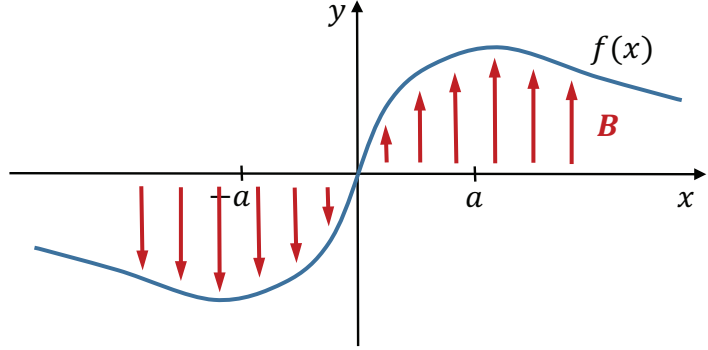}
\caption{Sketch of a general profile of the background magnetic field.}
\label{B_profile}
\end{minipage}
\end{center}
\end{figure}

To obtain the equations governing the tearing mode, we follow the standard procedure and represent the magnetic field as ${\bf B}(x,y)=B_0f(x){\hat {\bf y}}+{\bf b}(x,y)$, where $f(x)$ describes the profile of the reconnecting field, see Fig.~(\ref{B_profile}). Its typical scale, the thickness of the reconnection layer, is~$a$.
%\footnote{In the literature on tearing instability, the thickness of the layer is commonly denoted by $a$, while in the literature on turbulence it is denoted by $\lambda$.} 
The weak perturbation field can be represented through the magnetic potential ${\bf b}=-{\hat z}\times \nabla \psi=\left(\partial\psi/\partial y, -\partial\psi/\partial x \right)$. We assume that the background velocity is zero. The incompressible velocity perturbation is represented through the stream function ${\bf v}=\left(\partial \phi/\partial y, -\partial\phi/\partial x\right)$. We will neglect the effects of viscosity, but will keep the magnetic diffusivity. The magnetic induction and velocity momentum equations take the form e.g.,~\cite{biskamp2003}:
\begin{eqnarray}
\frac{\partial \psi}{\partial t}=\frac{\partial \phi}{\partial y} B_0f+\eta\nabla^2\psi, \\
\frac{\partial}{\partial t}\nabla^2\phi=B_0f\frac{\partial}{\partial y}\nabla^2\psi-B_0f''\frac{\partial\psi}{\partial y}.
\end{eqnarray}
We can use the Fourier transform in the $y$ direction, and represent the fluctuating fields as
\begin{eqnarray}
\psi={\tilde \psi}(\xi)\exp(ik_0 y)\exp(\gamma t), \\
\phi=-i{\tilde \phi}(\xi)\exp(ik_0 y)\exp(\gamma t).
\end{eqnarray}
In these expressions and in what follows we will use the dimensionless variables
\begin{eqnarray}
\xi=x/a,\label{xi}\\
{\tilde \eta}=\eta/\left(k_0V_Aa^2\right),\label{eta}\\
\lambda=\gamma/\left(k_0V_A\right),\label{lambda}\\
\epsilon=k_0a,\label{epsilon}
\end{eqnarray}
where $v_A$ is the Alfv\'en velocity associated with~$B_0$. The anisotropy parameter $\epsilon$ can be arbitrary. In applications to turbulence, however, the most relevant cases are those corresponding to $\epsilon\ll 1$, as only very anisotropic turbulent eddies become significantly affected by the tearing instability, e.g.,~\cite{loureiro2017,mallet2017,boldyrev_2017,loureiro2017a,mallet2017a,comisso2018,walker2018}. We, therefore, assume $\epsilon\ll 1$ in our discussion, which is the most difficult case to analyze. The results we derive can be extended to $\epsilon\approx 1$ if necessary.

In what follows we will use the tilded variables and omit the tilde sign. The dimensionless equations take a simple form:
\begin{eqnarray}
\lambda \psi=f\phi +\eta\left[\psi''-\epsilon^2\psi\right],\label{general_eq_1}\\
\lambda\left[\phi''-\epsilon^2\phi\right]=-f\left(\psi''-\epsilon^2\psi \right)+f''\psi,\label{general_eq_2}
\end{eqnarray}
where we denote by primes the derivatives with respect to $\xi$. An additional simplification of these equations can be obtained from the following consideration. We assume that $\eta$ and $\lambda$ are small parameters (this assumption can be verified a posteriori, from the obtained solution).  The range of scales where the terms including these parameters can be neglected will be called the {\em outer region}. The range of scales where they become significant will be called the {\em inner region}. 

We will solve equations (\ref{general_eq_1}) and (\ref{general_eq_2}) in the outer and inner regions separately,  and then asymptotically match the solutions. In the inner region, where as we will see, $\xi\ll 1$, we have $\partial^2/\partial \xi^2\gg 1$. Due to the smallness of $\eta$ and $\lambda$, the terms in the square brackets can be relevant only in the inner region, and, therefore, small $\epsilon^2$ terms can always be neglected in the square brackets. The equations then take the form
\begin{eqnarray}
\lambda \psi-f\phi =\eta \psi'',\label{eqs1} \\
-f\left(\psi''-\epsilon^2\psi \right)+f''\psi=\lambda\phi''.\label{eqs2}
\end{eqnarray}
Those are the equations describing the tearing instability in the very anisotropic case $\epsilon \ll 1$, and they are the main equations we are going to discuss in this work. The right-hand-side terms in these equations are relevant only in the inner region. In the outer region, they can be neglected.

\section{The outer equation}
We start with the outer region. We need to solve the equations
\begin{eqnarray}
\phi=\frac{\lambda}{f}\psi,\label{out2}\\
\psi''=\left(\frac{f''}{f}+\epsilon^2\right)\psi, \label{out1}
\end{eqnarray}
subject to the boundary conditions $\psi, \phi\to 0$ at $\xi\to\pm\infty$ (or to the periodic boundary conditions if the magnetic profile $f(\xi)$ is periodic). Before we discuss the solution we note that Eq.~(\ref{out1}) is the Schrodinger equation with zero energy. In general, it does not have a solution corresponding to given boundary conditions. So the solutions should be found separately for $\xi>0$ and $\xi<0$, but not in the region of small $\xi$ where this equation is not applicable. The solutions thus found will not, therefore, match smoothly, but will have a discontinuity in the derivative (a break) at $\xi=0$. 

We consider two exactly solvable model cases that correspond to particular profiles $f(\xi)$ of the reconnecting magnetic field. The first case is $f(\xi)=\tanh(\xi)$.  It is the so-called Harris profile \cite{harris_1962}. It corresponds to a magnetic field that value does not change in space except for a region of width $\xi\sim 1$, where it reverses its direction. The second solvable case is $f(\xi)=\sin(\xi)$~\cite{ottaviani1993}. In this case the magnetic field changes its strength and reverses its direction on the same scales. The latter case is arguably more relevant for the structures encountered in turbulence, and it is especially convenient for numerical studies as it allows for periodic boundary conditions. 

In the first case, $f=\tanh(\xi)$, the solution of Eq.~(\ref{out1}) for the magnetic field is (e.g., \cite{white1983}):
\begin{eqnarray}
\psi(\xi)=Ae^{-\epsilon \xi }\,\left[1+\frac{1}{\epsilon}\tanh(\xi)\right], \quad \xi\geq 0,\label{case1}\\
\psi(-\xi)=\psi(\xi).
\end{eqnarray}
The solution for the velocity function $\phi(\xi)$ is then easily found from Eq.~(\ref{out2}). In order to match with the inner solution, it is important to know the asymptotic forms of the velocity and magnetic fields for $\xi\ll 1$. Taking into account that $\epsilon \ll 1$, one obtains by expanding the $\tanh(\xi)$ in Eq.~(\ref{case1}) that
$\phi'(\xi)\sim -{A\lambda}/{\xi^2}+(2A\lambda \xi)/(3\epsilon)$. The second term can be neglected when $\xi^3\ll \epsilon$. The velocity $\phi'$ can then be formally expressed in this limit as  
\begin{eqnarray}
\phi'(\xi)\sim -\frac{\lambda}{\xi^2}\psi(0).\label{asympt_tanh}
\end{eqnarray}
As we will see later, in order to match the magnetic field one can define the tearing parameter 
\begin{eqnarray}
\Delta'=\frac{\psi'(\xi)-\psi'(-\xi)}{\psi(0)}, \quad \xi>0.
\end{eqnarray}
It is easy to see that in the region $\xi\ll 1$, the tearing parameter approaches a constant value 
\begin{eqnarray}
\Delta'= \frac{2}{\epsilon}.\label{delta_tanh}
\end{eqnarray}

In the second case, $f(\xi)=\sin(\xi)$, the solution periodic in $[-\pi, \pi]$ is:
\begin{eqnarray}
\psi(\xi)=A \sin\left[\sqrt{1-\epsilon^2} \left( \xi+\frac{\pi}{2\sqrt{1-\epsilon^2}}-\frac{\pi}{2}\right) \right], \quad \xi\geq 0,\\
\psi(-\xi)=\psi(\xi).
\end{eqnarray}
The derivative of the $\phi$ function is then found as $\phi'\sim -(A\lambda/\xi^2)(\pi\epsilon^2/4)+(A\lambda\xi/3)$. The second term can be neglected when $\xi^3\ll \epsilon^2$, in which case the $\phi'$ function has the asymptotic behavior that is formally identical to that obtained for the first case, 
\begin{eqnarray}
\phi'\sim -\frac{\lambda}{\xi^2}\psi(0).\label{asympt_sin}
\end{eqnarray}
Indeed, in this case $\psi(0)=A\pi\epsilon^2/4$. In the region $\xi\ll 1$, the tearing parameter approaches a constant value
\begin{eqnarray}
\Delta'= \frac{8}{\pi\epsilon^2}.\label{delta_sin}
\end{eqnarray}
Note the different scaling of this parameter with $\epsilon$ as compared to the previous result~(\ref{delta_tanh}).

It is important to check how deeply into the asymptotic region $\xi\ll 1$ the outer solutions can extend. In this region we estimate $f''\sim f \sim \xi$ for both tanh- and sine-shaped magnetic profiles. From Eq.~(\ref{out1}) we have $\psi''\sim \psi$. Then from Eq.~(\ref{out2}) we estimate $\phi''\sim (\lambda/\xi^3)\psi$. The right-hand sides in Eqs.~(\ref{eqs1}, \ref{eqs2}) are, therefore, small if 
\begin{eqnarray}
\eta\ll \lambda,\label{constr1}\\
\lambda^2\ll \xi^4. \label{constr2}
\end{eqnarray}

\section{The inner equation}
In the inner region we need to keep the right hand sides of Eqs.~(\ref{eqs1}) and~(\ref{eqs2}). For that the second derivatives of the fields should be large. For instance, we have to assume $\psi''\gg \psi$, which holds for $\xi\ll 1$. This allows us to simplify Eqs.~(\ref{eqs1},\ref{eqs2}) in the following way:
\begin{eqnarray}
\lambda \psi-\xi \phi=\eta \psi''\label{inner1},\\
-\xi \psi'' = \lambda\phi''. \label{inner2}
\end{eqnarray}
By differentiating Eq.~(\ref{inner1}) twice, we get $\lambda\psi''=\xi\phi''+2\phi'+\eta\psi''''$. We now exclude $\psi''$ and $\psi''''$ from this equation by using Eq.~(\ref{inner2}), $\psi''=-(\lambda/\xi) \phi''$. A few lines of algebra allow one to cast the resulting equation in the form
\begin{eqnarray}
\lambda^2\phi''+\left(\xi^2\phi' \right)'-\lambda\eta\left[\phi''''-2\left\{\frac{\phi''}{\xi} \right\}' \right]=0.
\end{eqnarray}
This equation can trivially be integrated once. Also, noting that it contains only derivatives of $\phi$, we may reduce the order by denoting $Y\equiv \phi'$. We then get:
\begin{eqnarray}
Y''-\frac{2}{\xi}Y'-\frac{1}{\eta\lambda}\left(\lambda^2+\xi^2\right)Y=C.\label{the_eq}
\end{eqnarray}
Before we analyze this equation further, we note that the first two terms in the left-hand side come from the resistive term in the induction equation. Also, one can directly verify that the term $\lambda^2$ in the parentheses would be absent if we used a common approximation treating $\psi(\xi)$ as a constant in the inner region, the so-called ``constant-$\psi$'' approximation. 

In order to find the constant of integration~$C$ we need to match asymptotically the solution of the inner Eq.~(\ref{the_eq}) with the solution of the outer equation. We know that the outer solution exists in $\xi^4\gg \lambda^2$, see Eq.~(\ref{constr2}), which, for $\xi<1$ also implies that $\xi^2\gg \lambda^2$. What asymptotic does the solution of Eq.~(\ref{the_eq}) have in region~(\ref{constr2})? There are two possibilities: $Y\sim \exp\left\{\xi^2/\sqrt{4\lambda\eta}\right\}$ and $Y\sim -C\eta\lambda/\xi^2 $. By evaluating different terms in Eq.~(\ref{the_eq}) for these asymptotic solutions, one can check that they hold for $\xi^4\gg \eta\lambda$, which is less restrictive than Eq.~(\ref{constr2}).

Obviously, the first asymptotic is not the solution we need, since the outer solution does not have an exponential growth at these scales. We, therefore, are interested in the inner solution with the asymptotic 
\begin{eqnarray} 
Y\sim -C\frac{\eta\lambda}{\xi^2}. \label{y_asympt}
\end{eqnarray}
In order to match this asymptotic expression with the outer solution $Y\equiv \phi'=-(\lambda/\xi^2)\psi(0)$, see expressions (\ref{asympt_tanh}) or (\ref{asympt_sin}), we need to choose $C=\psi(0)/\eta$. 

We note that the region where we asymptotically matched the two solutions is $\lambda^2\ll \xi^4\ll \epsilon^{4/3}$ in the case of the $\tanh$-profile, and $\lambda^2\ll \xi^4\ll \epsilon^{8/3}$ in the case of the sine-profile. Our solution, therefore, makes sense only when
\begin{eqnarray}
&\lambda^2/\epsilon^{4/3}\ll 1, \quad \mbox{for tanh-shaped profile},\\
&\lambda^2/\epsilon^{8/3}\ll 1, \quad \mbox{for sine-shaped profile},
\end{eqnarray}
which, as can be checked when the solution is obtained, are not restrictive conditions.

So far, we have matched the solutions for the velocity field, the $\phi'$ functions. To complete the asymptotic matching of the inner and outer solutions, we now need to match the magnetic fields, that is, the $\psi'$ function. This can be done in the following way. From Eq.~(\ref{inner2}) we get for the inner solution $\psi''=-(\lambda/\xi)\phi''$. Integrating this equation from $\xi\ll-\sqrt{\lambda}$ to $\xi \gg \sqrt{\lambda}$, which for the inner solution is equivalent to integrating from $-\infty$ to $\infty$, we obtain
\begin{eqnarray}
-\lambda\int\limits_{-\infty}^{+\infty}\frac{\phi''}{\xi}d\xi=\int\limits_{-\infty}^{+\infty}\psi''d\xi,
\end{eqnarray} 
which, recalling that $\phi'\equiv Y$,  can be rewritten as
\begin{eqnarray}
\int\limits_{-\infty}^{+\infty}\frac{Y^\prime}{\xi}d\xi=-\frac{\psi(0)}{\lambda}\Delta'.\label{condition}
\end{eqnarray}
This asymptotic matching condition will define the growth rate $\lambda$. 

It is convenient to change the variables in the following way. Let us introduce a function $G$ such that $Y=(\psi(0)/\lambda)G$, and the independent variable $\zeta=\xi^2/\lambda^2$. Then, the velocity-function equation Eq.~(\ref{the_eq}) takes the form
\begin{eqnarray}
4\zeta G''-2G'-\beta^2\left(1+\zeta\right)G=\beta^2,\label{the_equation}
\end{eqnarray}
where primes denote the derivatives with respect to $\zeta$, and we have denoted $\beta^2\equiv \lambda^3/\eta$. The matching condition (\ref{condition}) then takes the form 
\begin{eqnarray}
-2\int\limits_0^\infty \frac{1}{\sqrt{\zeta}}\frac{\partial G}{\partial\zeta}d\zeta=\lambda \Delta'.\label{boundary_condition}
\end{eqnarray}
We thus need to solve Eq.~(\ref{the_equation}), and then find the tearing growth rate $\lambda$ from the matching condition (\ref{boundary_condition}). 

The analytic theory of tearing mode is essentially based on Eqs.~(\ref{the_equation}) and (\ref{boundary_condition}). These equations can be solved exactly. Historically, however, a better known case is a simpler case of $\beta\ll 1$, the so-called FKR case~\cite{FKR}. The simplifying assumptions going into the FKR derivation are, however, easier to understand if one knows the exact solution of the problem. Here we, therefore, first concentrate on the exact solution.

\section{Solution of the inner equation}
Here we present the exact solution of the tearing equation~(\ref{the_equation}). This is an inhomogeneous equation, so its solution is a linear combination of a particular solution of the original inhomogeneous equation~(\ref{the_equation}) and a general solution of the homogeneous equation
\begin{eqnarray}
4\zeta {G_0}''-2{G_0}'-\beta^2\left(1+\zeta\right){G_0}=0.\label{the_homogeneous_equation}
\end{eqnarray}
The solution of the homogeneous equation~(\ref{the_homogeneous_equation}) has two possible asymptotics, $G_0\propto \exp{(\pm\beta\zeta/2)}$, at $\zeta\to\infty$. We obviously need to consider only the solutions behaving as~$G_0\propto \exp{(-\beta\zeta/2)}$. 

In order to figure out what this boundary condition means in terms of the original function $Y(\xi)$, we need to study the solutions of Eq.~(\ref{the_homogeneous_equation}) in more detail. As can be checked directly, at small $\zeta$ the solution has the following asymptotic behavior
\begin{eqnarray}
G_0(\zeta)\sim a_0\left(1-\frac{\beta^2}{2} \zeta +\frac{\beta^2}{4}\left\{1-\frac{\beta^2}{2}\right\}\zeta^2+\dots\right)+b_0\,\zeta^{3/2}\left( 1+\frac{3}{2}\zeta + \dots\right), \label{G_asympt}
\end{eqnarray}
where $a_0$ and $b_0$ are two arbitrary parameters. The ``$a_0$-part'' of the solution, which is a regular function at $\zeta=0$, corresponds to a solution of the homogeneous version of the original velocity equation (\ref{the_eq}) that is even in~$\xi$, while the singular ``$b_0$-part'' corresponds to a solution odd in~$\xi$.\footnote{This follows from the fact that the velocity function $Y(\xi)$ is analytic at $\xi=0$.} Equation (\ref{the_eq}) is symmetric with respect to $\xi\to-\xi$, therefore, each solution of (\ref{the_eq}) is a linear combinations of even and odd solutions. 

One can see from the asymptotic behavior~(\ref{G_asympt}) and from Eq.~(\ref{the_homogeneous_equation}) itself that in the odd solutions, the signs of the first and second derivatives are the same and they do not change on positive or negative axes. This means that every odd solution of the homogeneous version of equation (\ref{the_eq}) diverges at both $\xi\to \infty$ and $\xi\to -\infty$. In the case of even solutions, the same analysis can be applied to the function $G_0(\zeta)\exp\left(\beta^2\zeta/2\right)$ when $\beta<1$, from which it follows that every even solution of homogeneous Eq.~(\ref{the_eq}) diverges at both infinity limits as well.\footnote{Indeed, the function $P_0=G_0\exp\left(\beta^2\zeta/2\right)$ satisfies the equation $4\zeta P_0''=\left[2+4\zeta\beta^2\right]P_0'+\left[\beta^2-\beta^4\right]\zeta P_0$. Its even solution  has the asymptotic behavior $P_0\sim a_0+a_0\left(\beta^2/4\right)\left(1-\beta^2\right)\zeta^2$ at small $\zeta$. So, when $\beta<1$,  the function itself and, due to the equation it satisfies, its first and second derivatives have the same sign at $\zeta>0$. Therefore, the function $P_0$ diverges at infinity, which means that $G_0$ must diverge as well.} Only by choosing a particular relation between $a_0$ and $b_0$ can one cancel these divergences either at positive or negative infinity (but not at both). 

The method that we will use reproduces all the solutions that decline at $\zeta\to\infty$. Therefore, as we will see,  our derived expression for $G_0$ will contain both even and odd parts, but with a rigid relation between $a_0$ and $b_0$, to cancel this divergence.  We will need to remove such solutions since, as has been explained, they diverge either at $\xi\to-\infty$ or $\xi\to\infty$. Later, we will use this condition to uniquely define the solution.  

In order to find the general solution of Eq.~(\ref{the_equation}) we use the following method. The tearing equation~(\ref{the_equation}) is defined only for positive $\zeta$. We can, however, consider this equation on the whole $\zeta-$axis by formally extending the function $G$ to the negative values of the argument. For that we define
\begin{eqnarray}  
G(\zeta)=G_1(\zeta)\theta(\zeta)+G_2(\zeta)\left(1-\theta(\zeta)\right),\label{Gextended}
\end{eqnarray}
Where $G_1$ and $G_2$ are solutions of Eq.~(\ref{the_equation}) such that $G_1\to 0$ at $\zeta\to +\infty$, and $G_2\to 0$ at $\zeta\to -\infty$, and $\theta(\zeta)$ is the Heaviside step function. These solutions at positive and negative arguments are defined up to arbitrary, declining at infinity solutions of the homogeneous equation, therefore, they can always be chosen so that their amplitudes match at the origin, $G_1(0)=G_2(0)$. This provides a formal extension of the $G$ function to the negative arguments. Note that we match only the values of functions $G_1$ and $G_2$ at $\zeta=0$, but not their derivatives. 

If we consider the operator
\begin{eqnarray}
{\hat L}=4\zeta \frac{\partial^2}{\partial \zeta^2}-2\frac{\partial}{\partial \zeta}-\beta^2(1+\zeta),
\end{eqnarray}
we can directly verify that 
\begin{eqnarray}
{\hat L}G=\theta {\hat L}G_1+\left(1-\theta\right){\hat L}G_2=\beta^2.
\end{eqnarray}
Therefore, the extended function (\ref{Gextended}) satisfies the same equation~(\ref{the_equation}) on the entire real axis.  
This function declines at $\pm\infty$, therefore, we can Fourier transform equation~(\ref{the_equation}) using the standard definition
\begin{eqnarray}
G(\zeta)=\frac{1}{2\pi}\int\limits_{-\infty}^{\infty}G(k)e^{ik\zeta}dk.\label{fourier_transform}
\end{eqnarray}
In the Fourier space the equation takes the form
\begin{eqnarray}
\left(4k^2+\beta^2\right)G'+\left(10k-i\beta^2\right)G=2\pi i\beta^2\delta(k).\label{fourier}
\end{eqnarray}
The first-order ordinary differential equation~(\ref{fourier}) can easily be solved for $k<0$ and $k>0$. The solutions are
\begin{eqnarray}
G_{\pm}(k)=2\pi i {A}_{\pm}\left[1+\frac{4k^2}{\beta^2} \right]^{-\frac{5}{4}}\left[\frac{1+\frac{2ik}{\beta}}{1-\frac{2ik}{\beta}} \right]^{\frac{\beta}{4}},
\end{eqnarray}
where ${A}_{\pm}$ are two complex constants, and $\pm$ signs stand for the solutions defined on the positive and negative real $k$ axes, respectively. The function  $G(\zeta)$ is real, therefore ${A}_{-}=-{A}_{+}^*$. 

\begin{figure}[t]
\begin{center}
\begin{minipage}{4in}
\includegraphics[width=\columnwidth]{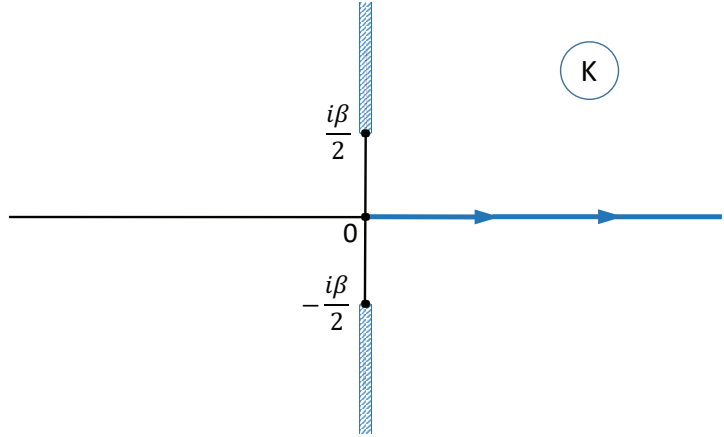}
\caption{Contour of integration in the $k$ plane in formula~(\ref{Gintegral1}). Possible branch cuts necessary to define analytic continuations of the integrands in~(\ref{Gintegral1}) are also shown.}
\label{k_contour}
\end{minipage}
\end{center}
\end{figure}

The delta-function in the right-hand side of Eq.~(\ref{fourier}) implies that the function $G(k)$ is discontinuous on the real $k$ axis at $k=0$, with the discontinuity condition ${A}_+-{A}_-=1$. In what follows we will simply denote  ${A}_+=A$  and ${A}_-=-A^*$, so that the discontinuity condition reads $A+A^*=1$.  As a result, the function $G$ can be represented as:
\begin{eqnarray}
G(\zeta)=i A\int\limits_{0}^{+\infty}\left[1+\frac{4k^2}{\beta^2} \right]^{-\frac{5}{4}}\left[\frac{1+\frac{2ik}{\beta}}{1-\frac{2ik}{\beta}} \right]^{\frac{\beta}{4}}e^{ik\zeta}\,dk 
-i A^*\int\limits_{0}^{+\infty}\left[1+\frac{4k^2}{\beta^2} \right]^{-\frac{5}{4}}\left[\frac{1-\frac{2ik}{\beta}}{1+\frac{2ik}{\beta}} \right]^{\frac{\beta}{4}}e^{-ik\zeta}\,dk.\,\,\label{Gintegral1}
\end{eqnarray}
In these integrals the branches of the integrands must be chosen so that they coincide at $k=0$, and the integration is performed along the positive real line in a complex plane, see Fig.(\ref{k_contour}). We note that the discontinuity condition defines only the real part of the complex coefficient~$A$, but leaves its imaginary part arbitrary. This reflects the fact the solution is not defined uniquely, but only up to an arbitrary solution of the homogeneous equation~(\ref{the_homogeneous_equation}).\footnote{It is easy to see that the solution of the homogeneous equation is given by the integral
\begin{eqnarray}
G_0(\zeta)=A_0\int\limits_{-\infty}^{+\infty}\left[1+\frac{4k^2}{\beta^2} \right]^{-\frac{5}{4}}\left[\frac{1+\frac{2ik}{\beta}}{1-\frac{2ik}{\beta}} \right]^{\frac{\beta}{4}}\exp(ik\zeta)\,dk,
\end{eqnarray}
where $A_0$ is an arbitrary real constant.}  

For practical calculations, it is convenient to modify Eq.~(\ref{Gintegral1}) further. In the first integral of Eq.~(\ref{Gintegral1}) we change the variable of integration
\begin{eqnarray}
k=\left(\frac{\beta}{2i}\right)\frac{q-1}{q+1},\label{subst1}
\end{eqnarray}
while in the second integral we choose
\begin{eqnarray}
k=-\left(\frac{\beta}{2i}\right)\frac{p-1}{p+1}.\label{subst2}
\end{eqnarray}
Expression (\ref{Gintegral1}) now takes the form
\begin{eqnarray}
G(\zeta)=-A\frac{\beta}{4\sqrt{2}}\int\limits_{-1}^{1}\frac{(q+1)^{\frac{1}{2}}}{q^{\frac{5}{4}}}q^{\frac{\beta}{4}}e^{\frac{\beta}{2}\frac{q-1}{q+1}\zeta}\,dq -A^*\frac{\beta}{4\sqrt{2}}\int\limits_{-1}^{1}\frac{(p+1)^{\frac{1}{2}}}{p^{\frac{5}{4}}}p^{\frac{\beta}{4}}e^{\frac{\beta}{2}\frac{p-1}{p+1}\zeta}\,dp.\label{Gintegral2}
\end{eqnarray}

\begin{figure}[ht]
\begin{center}
\begin{minipage}{4in}
\includegraphics[width=\columnwidth]{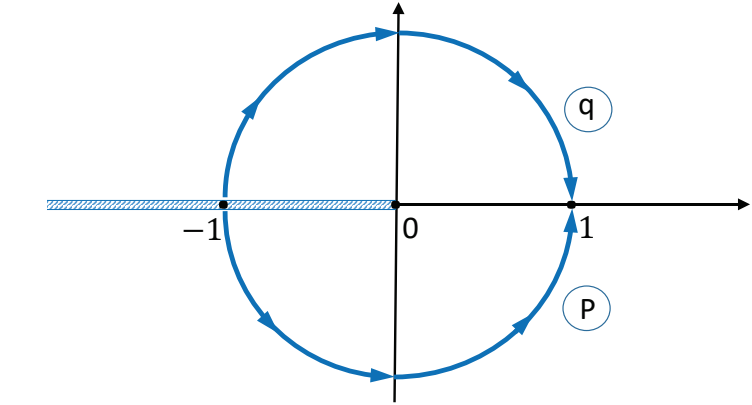}
\caption{Contours of integration in the $q$ and $p$ complex planes in formula~(\ref{Gintegral2}).}
\label{qp_contour}
\end{minipage}
\end{center}
\end{figure}

The integrals in Eq.~(\ref{Gintegral2}) look identical. However, they differ by the contours of integration that follow from the changes of variables~(\ref{subst1}) and~(\ref{subst2}). If in each integral of Eq.~(\ref{Gintegral1}), the contours of integration  lie along the real axis in the complex $k$ plane, see Fig.~(\ref{k_contour}), then the corresponding contours in the $q$ and $p$ planes are defined as shown in Fig~(\ref{qp_contour}).  

\begin{figure}[ht]
\begin{center}
\begin{minipage}{4in}
\includegraphics[width=\columnwidth]{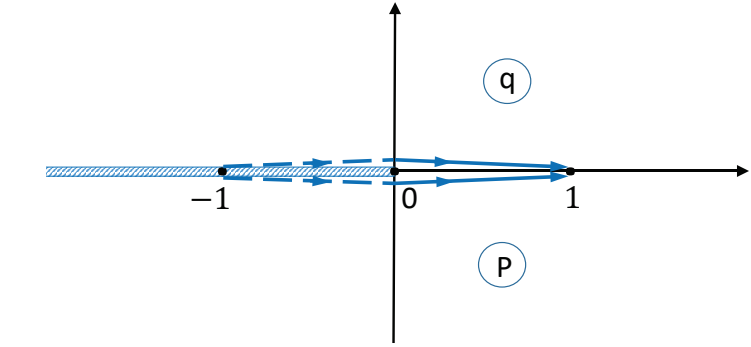}
\caption{Equivalent contours of integration in the $q$ and $p$ complex planes in formula~(\ref{Gintegral3}).}
\label{qp_contour_new}
\end{minipage}
\end{center}
\end{figure}
It is easy to see that for $\zeta>0$, these integrals will not change if we deform the contours to coincide with the real axis, as shown in Fig~(\ref{qp_contour_new}). This way, one of the contours of integration has to go above the branch cut, and the other one below. It is also useful to integrate  by parts once, in order to avoid dealing with too strong a singularity at the origin,
\begin{eqnarray}
G(\zeta)=\frac{A \beta}{1-\beta}\left[1-\frac{1}{\sqrt{2}}\int\limits_{-1}^1q^{\frac{\beta-1}{4}}\frac{d}{dq}\left\{ \left(q+1\right)^{\frac{1}{2}}e^{\frac{\beta}{2}\zeta\frac{q-1}{q+1}}\right\}dq\right]\nonumber \\
+\frac{A^* \beta}{1-\beta}\left[1-\frac{1}{\sqrt{2}}\int\limits_{-1}^1q^{\frac{\beta-1}{4}}\frac{d}{dq}\left\{ \left(q+1\right)^{\frac{1}{2}}e^{\frac{\beta}{2}\zeta\frac{q-1}{q+1}}\right\}dq\right].
\label{Gintegral3}
\end{eqnarray}

In the interval $(0, 1]$ both integrals are the same, and their sum can be simplified since $A+A^*=1$.  In the interval $[-1, 0)$, however, the integrals have different phases. The integrals over the dashed lines add up to $A\exp\left[i\pi(\beta-1)/4\right]+A^*\exp\left[-i\pi(\beta-1)/4\right]$.  
Since the imaginary part of $A$ is arbitrary, this sum is arbitrary as well (we assume $\beta<1$).
%We can separate the real and imaginary parts in their amplitudes, $A=(1+i\alpha)/2$, where $\alpha$ is an arbitrary real parameter. The sum of the integrals over the dashed lines will then lead to an overall coefficient 
%\begin{eqnarray}
%&A_0\equiv A\exp\left[i\pi(\beta-1)/4\right]+A^*\exp\left[-i\pi(\beta-1)/4\right]=\nonumber \\
%&\cos\left[\pi(\beta-1)/4\right] -\alpha \sin\left[\pi(\beta-1)/4\right].\quad
%\end{eqnarray}
As a result the solution for $\zeta>0$ can be represented as
\begin{eqnarray}
G(\zeta)=\frac{\beta}{1-\beta}\left[1-\frac{1}{\sqrt{2}}\int\limits_0^1q^{\frac{\beta-1}{4}}\frac{d}{dq}\left\{ \left(q+1\right)^{\frac{1}{2}}e^{\frac{\beta}{2}\zeta\frac{q-1}{q+1}}\right\}dq\right]\nonumber \\
%&-\frac{A_0\beta}{\sqrt{2}(1-\beta)}
+\, C_0\int\limits_{-1}^0 |q|^{\frac{\beta-1}{4}}\frac{d}{dq}\left\{ \left(q+1\right)^{\frac{1}{2}}e^{\frac{\beta}{2}\zeta\frac{q-1}{q+1}}\right\}dq, \label{Gintegral4}
\end{eqnarray}
where $C_0$ is an arbitrary parameter. The first term in this expression is a particular solution of the inner equation (\ref{the_equation}). It has been originally derived in \cite{coppi1966,coppi_resistive_1976,abc1978}, where a different approach involving expansion in the Laguerre polynomials, has been used. The second term, including a free parameter $C_0$, represents the solution of the homogeneous equation~(\ref{the_homogeneous_equation}), the zero mode. 

It is easy to see that the particular solution is an analytic function at $\zeta=0$, so it describes an even solution of Eq.~(\ref{the_eq}). Also, this solution converges at~$\zeta\to\infty$. Therefore, it represents a solution of Eq.~(\ref{the_eq}) that converges at $\xi\to \pm\infty$. The zero mode, on the contrary, is non-analytic at $\zeta=0$, since its second derivative diverges there. It is therefore a combination of odd and even solutions of homogeneous equation~(\ref{the_eq}). According to what was said in the beginning of this section, a solution of the homogeneous equation diverges at either $\xi \to -\infty$ or $\xi \to +\infty$. We thus have to require $C_0=0$, which removes these divergences. The zero mode is therefore absent and  the solution is given by the first term in expression~(\ref{Gintegral4}).

\section{Tearing rate in the limit $\beta\ll 1$ (the FKR case)}
Now that we have obtained the general solution for the inner region, we can find the tearing mode growth rate by substituting this solution into the matching condition~(\ref{boundary_condition}). 
Before considering the general case, however, we discuss the important limit of $\beta\ll 1$, the so-called FKR case \cite{FKR} (we recall that $\beta^2=\lambda^3/\eta$). In this limit, we can approximate $q^{(\beta-1)/4}\approx q^{-1/4}$ in Eq.~(\ref{Gintegral4}). It is possible to show that this approximation is equivalent to the ``constant-$\psi$'' approximation discussed after Eq.~(\ref{the_eq}), which demonstrates the equivalence of the constant-$\psi$ approximation to the FKR case.  We see that in this case the integral depends on $\zeta$ only through the combination $\beta\zeta$. The matching condition (\ref{boundary_condition}) now reads
\begin{eqnarray}
C\beta^{3/2}=\lambda\Delta',\label{bc_fkr}
\end{eqnarray}  
where the constant $C$ is given by the integral
\begin{eqnarray}
 C=\int\limits_0^{\infty} \sqrt{\frac{2}{x}}\frac{d}{dx}\left[\int\limits_0^1q^{-\frac{1}{4}}\frac{d}{dq}\left\{ \left(q+1\right)^{\frac{1}{2}}e^{\frac{x}{2}\frac{q-1}{q+1}}\right\}dq\right]dx.\quad\quad
\end{eqnarray}
One can easily do this integral by changing the order of integrations. The answer is expressed through the gamma functions, 
\begin{eqnarray}
C=\left(\frac{\pi}{2}\right)\frac{\Gamma\left(\frac{3}{4}\right)}{\Gamma\left(\frac{5}{4}\right)}\approx 1.45.\label{c_fkr}
\end{eqnarray}

From Eq.~(\ref{bc_fkr}) we find the growth rate of the tearing mode in the FKR regime of $\beta\ll 1$, 
\begin{eqnarray}
\lambda=C^{-4/5}\eta^{3/5}{\Delta'}^{\,4/5}.\label{lambda_fkr}
\end{eqnarray}
In the {\em dimensional} form, this expression can be rewritten as:
\begin{eqnarray}
\gamma=2^{4/5}C^{-4/5}\eta^{3/5}k_0^{-2/5}V_A^{2/5}a^{-2}\label{gamma_tanh}
\end{eqnarray}
 for the $\tanh$-shaped magnetic profile, and 
\begin{eqnarray}
\gamma=8^{4/5}\pi^{-4/5}C^{-4/5}\eta^{3/5}k_0^{-6/5}V_A^{2/5}a^{-14/5}\label{gamma_sin}
\end{eqnarray}
for the sine-shaped magnetic profile. In order to obtain these results we have substituted the expressions (\ref{delta_tanh}) and (\ref{delta_sin}) for the corresponding parameters $\Delta'$.

Two important points should be made about the FKR solution. First, since in this limit the inner function $G(\zeta)$ depends on scale only through the combination $\beta\zeta$, this function has a characteristic length scale, $\zeta=1/\beta$, which is the so-called {\em inner scale}. In terms of the $\xi$ variable, this scale is $\xi=(\lambda\eta)^{1/4}\gg \lambda$. In dimensional form the inner scale $x=\delta$ is given by
\begin{eqnarray}
\delta=\left(\frac{\gamma\eta a^2}{k_0^2 V_A^2} \right)^{1/4},\label{inner_scale}
\end{eqnarray} 
where the growth rate $\gamma$ is given by either (\ref{gamma_tanh}) or (\ref{gamma_sin}) depending on the chosen magnetic profile.

Second, we see that the growth rate and the inner scale formally diverge for $k_0\to 0$. This is obviously an unphysical behavior. It reflects the fact that the approximation $\beta\ll 1$ is not valid in this case. We will now present the solution for the growth rate in the general case, where the inner function $G(\zeta)$ is given by the exact expression~(\ref{Gintegral4}). We will see that the growth rate does not diverge, but reaches a maximal value at a certain small value of~$k_0a$.

\section{Tearing rate in the general case}
Here we analyze the general case, the so-called Coppi case \cite{coppi1966,coppi_resistive_1976,abc1978}. We need to evaluate the integral in the left-hand-side of~(\ref{boundary_condition}) using the exact expression for the $G$ function given by~(\ref{Gintegral4}).  This can be easily done in the same way as we obtained~(\ref{c_fkr}),
\begin{eqnarray}
-2\int\limits_0^\infty \frac{1}{\sqrt{\zeta}}\frac{\partial G}{\partial\zeta}d\zeta=-\frac{\pi}{8}\beta^{3/2}\frac{\Gamma\left(\frac{\beta-1}{4}\right)}{\Gamma\left(\frac{\beta+5}{4}\right)}.
\end{eqnarray}
The growth rate is found from the transcendental equation \cite{coppi_resistive_1976}:
\begin{eqnarray}
-\frac{\pi}{8}\beta^{3/2}\frac{\Gamma\left(\frac{\beta-1}{4}\right)}{\Gamma\left(\frac{\beta+5}{4}\right)}=\lambda\Delta'.\label{general_condition}
\end{eqnarray}
We see that the left-hand-side of this expression is positive when $\beta<1$, and, therefore, the instability is possible only in this case. In the limit of $\beta\ll 1$, we recover the results discussed in the previous section. The left-hand-side of Eq.~(\ref{general_condition}) is small in this limit. The low-$\beta$ approximation, however, breaks down at $k_0\to 0$. As we will see momentarily, in this limit the solution corresponds to $\beta\to 1$. Indeed, equation~(\ref{general_condition}) can be approximated in this case as
\begin{eqnarray}
\sqrt{\pi}\frac{\beta^{3/2}}{1-\beta}=\lambda\Delta'.
\end{eqnarray}
Recalling now that $\beta^2\equiv \lambda^3/\eta$, we arrive at the equation for the growth rate
\begin{eqnarray}
\sqrt{\pi}\frac{\beta^{5/6}}{1-\beta}=\eta^{1/3}\Delta'.
\end{eqnarray}
From the definitions of the dimensionless parameters $\eta$ and $\Delta'$, we see that as $k_0a\to 0$, the right-hand-side of this equation diverges. This is possible only if $\beta\to 1$ for such solutions. We, therefore, see that $\lambda^3=\eta$ is the equation that defines the tearing growth rate in this case. In the {\em dimensional} units, this equation gives
\begin{eqnarray}
\gamma=\eta^{1/3}V_A^{2/3}k_0^{2/3}a^{-2/3},\label{gamma_coppi}
\end{eqnarray}
which is termed the Coppi solution in \cite{loureiro_instability_2007}. The remarkable fact is that as $k_0$ decreases, the Coppi growth rate decreases as well. This is opposite to the behavior of the FKR growth rate that increases at decreasing~$k_0$.  This means that there must exist a maximal growth rate of the tearing instability, $\gamma_*$, attainable at a certain wave number ${k_0}_*$.  One can {\em define} this critical wave number as  the wave number at which the Coppi growth rate~(\ref{gamma_coppi}) formally matches the FKR growth rate~(\ref{gamma_tanh}) for the $\tanh$-shaped magnetic profile or~(\ref{gamma_sin}) for the sine-shaped profile.  

For the tanh-shaped magnetic profile, a simple algebra gives for the critical wavenumber and the corresponding maximal tearing growth rate:
\begin{eqnarray}
{k_0}_*=(2/C)^{3/4}\eta^{1/4}V_A^{-1/4}a^{-5/4},\\
\gamma_*=(2/C)^{1/2}\eta^{1/2}V_A^{1/2}a^{-3/2}.\label{gamma_tanh_coppi}
\end{eqnarray}
For the sine-shaped profile, the answer is:
\begin{eqnarray}
{k_0}_*=(8/\pi C)^{3/7}\eta^{1/7}V_A^{-1/7}a^{-8/7}, \\
\gamma_*=(8/\pi C)^{2/7}\eta^{3/7}V_A^{4/7}a^{-10/7}.\label{gamma_sin_coppi}
\end{eqnarray}
In the general case, the inner solution $G(\zeta)$ is not a universal function depending only on $\beta\zeta$. However, for $\beta\approx 1$, this function approaches its asymptotic behavior $G(\zeta)\sim 1/\zeta$ at $\zeta\gg 1$. The typical scale (the inner scale) of this solution is therefore $\zeta=1$, which in terms of the $\xi$ variable reads $\xi=\lambda$. In dimensional variables, the inner scale in this case is 
\begin{eqnarray}
\delta=\frac{\gamma a}{k_0v_A},
\end{eqnarray} 
where $\gamma$ is given by Eq.~(\ref{gamma_coppi}), or by expressions (\ref{gamma_tanh_coppi}) or (\ref{gamma_sin_coppi}) for the fastest growing modes.

\section{Tearing rates in the presence of a shear flow}
\label{shear_flow}
In turbulent systems, magnetic field fluctuations are accompanied by velocity fluctuations, so that both  magnetic and velocity shears are present in a turbulent eddy, e.g.,~\cite{boldyrev_spectrum_2006}. It is therefore useful to comment on the influence of a velocity shear on the tearing instability. A velocity shear across the current layer is expected to be less intense than the magnetic shear, otherwise, such a layer would be destroyed by the Kelvin-Helmholtz instability, e.g.,~\cite{biskamp2003,walker2018}. In MHD turbulence, fluctuations indeed have excess of the magnetic energy over the kinetic energy; the difference between the kinetic and magnetic energies, the so-called residual energy, is negative in the inertial interval, e.g., \cite{boldyrev_res2011,wang_res2011,boldyrev_res2012,boldyrev_res2012a,chen_res2013}.  

We assume that, similarly to the magnetic field, the background velocity has the structure ${\bf v}_0(x,y)=v_0(x){\hat {\bf y}}$. In the presence of this velocity field, the dimensionless system of equations~(\ref{general_eq_1}), (\ref{general_eq_2}) becomes
\begin{eqnarray}
\lambda \psi= f\phi +\eta\left[\psi''-\epsilon^2\psi\right]-i{\tilde v}_0\psi,\label{v_general_eq_1}\\
\lambda\left[\phi''-\epsilon^2\phi\right]=-f\left(\psi''-\epsilon^2\psi \right)+f''\psi -i{\tilde v}_0\left(\phi''-\epsilon^2\phi\right)+i{\tilde v}^{\prime\prime}_0\phi,\label{v_general_eq_2}
\end{eqnarray}
where ${\tilde v}_0=v_0(x)/V_A$ is the background velocity profile normalized by the Alfv\'en speed associated with the magnetic profile. Similarly to equations~(\ref{general_eq_1}), (\ref{general_eq_2}), the modified equations can be simplified in the case of small $\eta$ and $\lambda$ as:
\begin{eqnarray}
\lambda \psi= f\phi +\eta\psi''-i{\tilde v}_0\psi,\label{v_eqs_1}\\
\lambda\phi''=-f\left(\psi''-\epsilon^2\psi \right)+f''\psi -i{\tilde v}_0\left(\phi''-\epsilon^2\phi\right)+i{\tilde v}^{\prime\prime}_0\phi.\label{v_eqs_2}
\end{eqnarray}

In the {\em outer region}, we have from Eqs.~(\ref{v_eqs_1}), (\ref{v_eqs_2}):
\begin{eqnarray}
\lambda \psi= f\phi -i{\tilde v}_0\psi,\label{v_eqs_1a}\\
-f\left(\psi''-\epsilon^2\psi \right)+f''\psi -i{\tilde v}_0\left(\phi''-\epsilon^2\phi\right)+i{\tilde v}^{\prime\prime}_0\phi=0.\label{v_eqs_2a}
\end{eqnarray} 
A general analysis of the problem is, unfortunately, not very transparent, e.g.,~\cite{chen_morrison1990}. A simplified but quite informative treatment is, however, possible when the velocity profile is similar to that of the magnetic field~\cite{hofmann1975}. We therefore assume that ${\tilde v}_0(x)=\alpha f(x)$, where $-1<\alpha <1$.  
%see Fig.~(\ref{BV_profile}). 

We see from Eq.~(\ref{v_eqs_1a}) that the shear velocity introduces a Doppler shift that competes with the growth rate (${\tilde v}(\xi)/\lambda=k_0v(\xi)/\gamma$ in dimensional units). Since in the outer region $v\lesssim V_A$, the Doppler shift dominates, and we can neglect the term containing~$\lambda$ in Eq.~(\ref{v_eqs_1a}).  Expressing $\phi$ from Eq.~(\ref{v_eqs_1a}), and substituting it into Eq.~(\ref{v_eqs_2a}) one obtains after simple algebra~\cite{hofmann1975}:
\begin{eqnarray}
\left(1-\alpha^2\right)\left\{\psi''-\epsilon^2\psi-\frac{f''}{f}\psi \right\}=0. 
\end{eqnarray}
Since $\alpha^2\neq 1$, the magnetic-field outer equation in this case is identical to the outer equation without a velocity shear~(\ref{out2}). The asymptotic behavior of the outer solution for the velocity field is, however, different from Eqs.~(\ref{asympt_tanh}, \ref{asympt_sin}), and it can be found from Eq.~(\ref{v_eqs_1a}) to the zeroth order in $\lambda/{\tilde v}_0$:
\begin{eqnarray}
\phi (\xi)\sim i\alpha \psi(\xi), \quad \xi\ll 1.\label{v_phi_asympt}
\end{eqnarray}
This expression holds in the asymptotic matching region $\delta \ll \xi \ll 1$, where $\delta$ is the inner scale.

In the {\em inner region} $\delta\sim \xi\ll 1$ we have
\begin{eqnarray}
\lambda\psi +i\alpha\xi\psi=\xi\phi+\eta\psi'' ,\label{v_inner_1}\\
\lambda\phi''+i\alpha\xi\phi''=-\xi\psi'' .\label{v_inner_2}
\end{eqnarray}
In order to match the inner solution for the velocity field with the outer solution, we derive from Eq.~(\ref{v_inner_2}):
\begin{eqnarray}
\lambda \int\limits^{+\infty}_{-\infty}\frac{\phi''}{\xi}d\xi=-i\alpha\phi'\vert^{+}_{-}-\psi'\vert^{+}_{-},
\end{eqnarray}
where the integral goes from $\xi\ll -\delta$ to $\xi\gg \delta$, and we denote by $\vert^+_-$ the jumps of the quantities across the inner layer between the indicated limits. We can use the asymptotic form~(\ref{v_phi_asympt}) to evaluate the jump of $\phi'$. From Eq.~(\ref{v_phi_asympt}) we get
\begin{eqnarray}
\lambda \int\limits^{+\infty}_{-\infty}\frac{\phi''}{\xi}d\xi \,\sim \, -\left(1-\alpha^2\right)\psi(0)\Delta'.\label{matching}
\end{eqnarray}
In order to do the integral in Eq.~(\ref{matching}) we need to know the velocity function $\phi(\xi)$ in the inner region. To the best of our knowledge, the exact solution of the inner equations (\ref{v_inner_1}, \ref{v_inner_2}) is not available. We may, however, estimate the integral in the left-hand-side of Eq.~(\ref{matching}) in the following way. We note that if $\lambda\gg \alpha\delta$, then the shear flow does not affect the inner region of the tearing mode. The inner solution (and the resulting scaling of the growth rate with the Lundquist number~$S=aV_A/\eta$ and the anisotropy parameter~$k_0a$)  can, therefore, be qualitatively affected by the shear flow only if $\lambda \ll \alpha \delta$, which is essentially the FKR limit. We thus assume this limit in what follows.

Similarly to the case without a flow, the shear-modified FKR limit implies the ``constant-$\psi$'' approximation, which reads to the zeroth order in the small parameter $\lambda/(\alpha\delta)$: $\phi(\xi)\sim i\alpha\psi(\xi)\sim$~const at $\xi\ll 1$. This solution trivially satisfies Eqs.~(\ref{v_inner_1}, \ref{v_inner_2})  to the zeroth order. Obviously, the zeroth order solution does not contribute to the integral in (\ref{matching}), so in order to evaluate this integral we need to go to the first order in $\lambda/(\alpha\delta)$. From Eq.~(\ref{v_inner_2}) we estimate $\psi''\sim -i\alpha \phi''$ and substituting this into Eq.~(\ref{v_inner_1}), we get
\begin{eqnarray}
\lambda\psi(0) +\alpha^2\xi\phi=\xi\phi-i{\eta}{\alpha}\phi'' ,\label{b_inner_1}
\end{eqnarray}
where in the first term in the left-hand-side we have substituted the zeroth order solution. In the region $\xi \gg \delta$, the resistive term is not important, and we obtain the asymptotic form for the first-order velocity field:
\begin{eqnarray}
\phi\sim \frac{\lambda\psi(0)}{(1-\alpha^2)\xi}.\label{v_inner_asympt}
\end{eqnarray}
This expression does contribute to the integral in Eq.~(\ref{matching}). It diverges as $\xi$ decreases, until $\xi$ becomes as small as $\delta$, at which scale the resistive term becomes important and $\phi$ does not grow anymore. We may then estimate the integral in Eq.~(\ref{matching}) as 
\begin{eqnarray}
\lambda\int\limits^{+\infty}_{-\infty}\frac{\phi''}{\xi}d\xi\,\,\sim \,\, 2\lambda\int\limits^{+\infty}_{\delta}\frac{\phi''}{\xi}d\xi
\,\,\sim \,\, -\frac{\lambda^2\psi(0)}{(1-\alpha^2)\delta^3}\,,\label{estimate}
\end{eqnarray}
where the inner scale $\delta$ is, in turn, estimated from balancing the resistive term in Eq.~(\ref{b_inner_1}) with the other terms: $\delta^3\sim \alpha\eta/(1-\alpha^2)$.  Substituting this into Eq.~(\ref{estimate}) and then into Eq.~(\ref{matching}) we finally obtain
\begin{eqnarray}
\lambda \sim \left[\eta\,|\alpha | \left(1-\alpha^2\right)\Delta'\right]^{1/2}.
\end{eqnarray}  
This result coincides with the more detailed derivations performed in \cite{hofmann1975,chen_morrison1990} up to a numerical coefficient of order unity. 
\begin{figure}[t]
\begin{center}
\begin{minipage}{4in}
\includegraphics[width=\columnwidth]{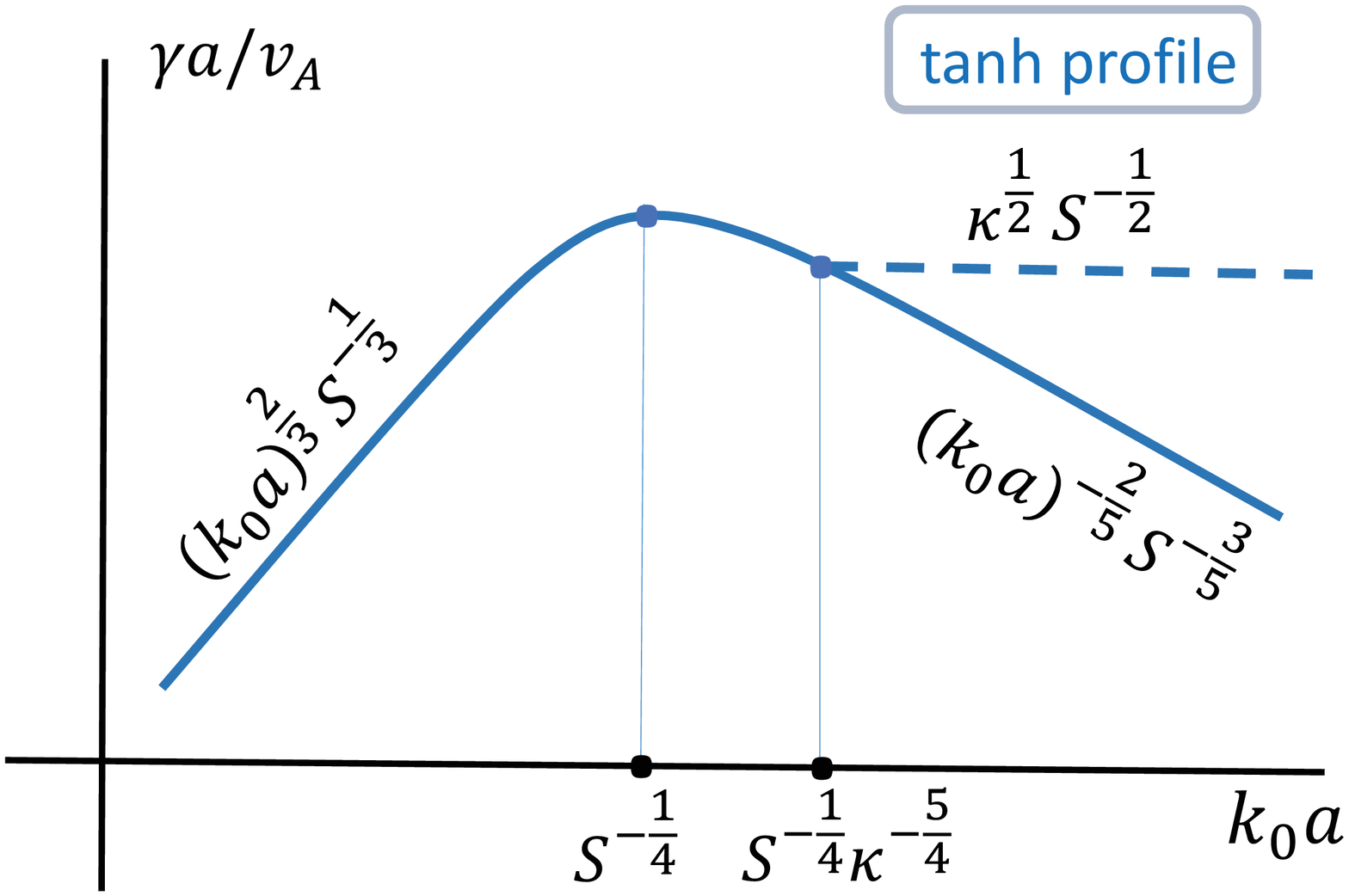}
\includegraphics[width=\columnwidth]{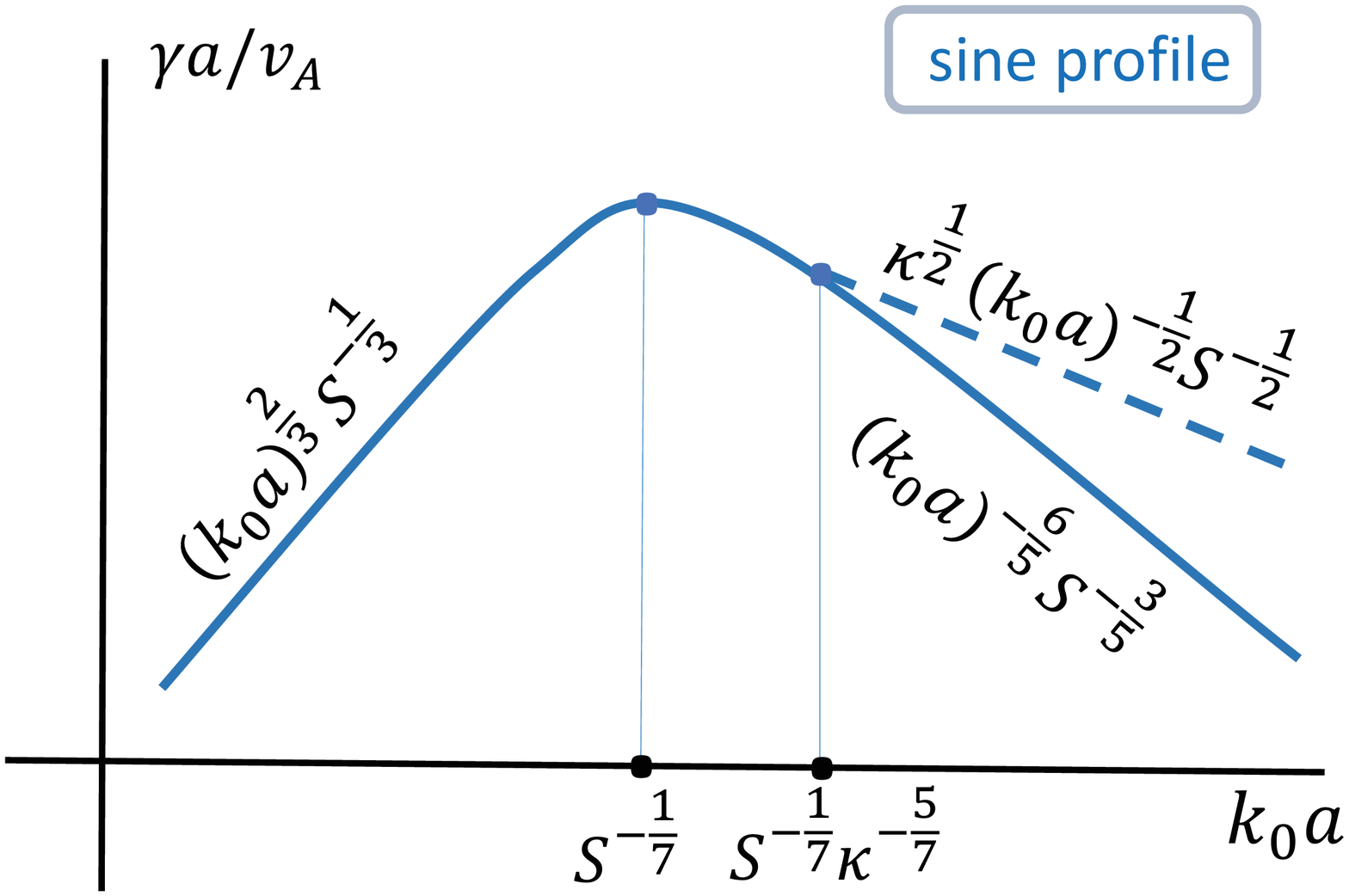}
\caption{Sketch of the tearing-mode growth rate as a function of $(k_0a)$ for the tanh-like magnetic profile (upper panel) and the sine-like profile (lower panel). The solid lines correspond to the growth rates without a shear flow. The dashed lines show the growth rates in the shear-modified FKR regimes. Here $S=aV_A/\eta$ and $\kappa=|\alpha|\left(1-\alpha^2 \right)<1$.}
\label{v_gammas}
\end{minipage}
\end{center}
\end{figure}

In {\em dimensional} variables, this growth rate takes the following form for the tanh-like magnetic profile: 
\begin{eqnarray}
\gamma\sim \kappa^{1/2}\eta^{1/2}V_A^{1/2}a^{-3/2},\label{v_gamma_tanh}
\end{eqnarray}
while for the sine-like profile we obtain
\begin{eqnarray}
\gamma\sim \kappa^{1/2}\eta^{1/2}V_A^{1/2}a^{-2}k_0^{-1/2},
\end{eqnarray}
where we denote $\kappa\equiv |\alpha | \left(1-\alpha^2\right)< 1$. These solutions are shown in Fig.~(\ref{v_gammas}), together with the expressions (\ref{gamma_tanh}), (\ref{gamma_sin}), and (\ref{gamma_coppi}), corresponding to the cases without a flow. We see that a shear flow changes the scaling of the growth rate in the FKR regime, but not in the Coppi regime. In particular, it does not affect the scaling of the fastest growing mode.

Note that the growth rate (\ref{v_gamma_tanh}) corresponding to the tanh-like profile is degenerate in that it is independent of the anisotropy parameter~$k_0 a$. This explains why the early analysis of \cite{carbone1990} based on the Iroshnikov-Kraichnan theory of MHD turbulence, which assumes isotropy of the turbulent fluctuations (i.e, it implies $k_0a\sim$~const), formally led to the same scaling of the fastest growing mode, $\gamma_* \sim S^{-1/2}$, as the analysis of \cite{loureiro2017,mallet2017} for the tanh-like magnetic profile, cf. Eq.~(\ref{gamma_tanh}). In the non-degenerate sine-like case, however, the fastest growing rate scales as $\gamma_*\sim S^{-3/7}$ cf.~Eq.~(\ref{gamma_sin}), which is different from $\gamma_* \sim S^{-1/2}$ assumed in~\cite{carbone1990}.

\section{Conclusions}
We have reviewed the derivation of the anisotropic tearing mode by considering in detail two solvable cases corresponding to the tanh-shaped and sine-shaped magnetic shear profiles. Given large enough anisotropy, the dominating tearing mode has the dimension $\sim 1/{k_0}_*$ and grows with the rate $\gamma_*$. We see that these parameters depend on the assumed magnetic shear profile, and they are not universal. Their derivation requires one to go beyond the simplified FKR model generally covered in textbooks. We have presented an effective method for solving the inner equation for the current layer in the general case. We have also discussed the influence on the tearing instability of shearing flows that typically accompany magnetic profiles generated by turbulence.  Our work provides a detailed and self-contained discussion of the methods required for the study of tearing effects in turbulent systems.

\ack
SB was partly supported by the NSF grant no. PHY-1707272, NASA Grant No. 80NSSC18K0646, and by the Vilas Associates Award from the University of Wisconsin - Madison.
NFL was supported by the NSF-DOE Partnership in Basic Plasma Science and Engineering, award no. DE-SC0016215 and by the NSF CAREER award no. 1654168.

\section*{References}

%-----------------------------------------------------------------------------------
%\bibliographystyle{apj}

%\bibliography{master_calculations}

%=Control: year (1) truncated

\providecommand{\newblock}{}

\end{document}